\documentclass[prb,aps,twocolumn]{revtex4}
\usepackage{color,graphicx,bm,amsmath,dsfont}

\begin{document}

\newcommand{\half}{\ensuremath{\frac{1}{2}}}
\newcommand{\degree}{\ensuremath{^{\circ}}}
\newcommand{\adag}{\ensuremath{a^{\dagger}}}
\newcommand{\bra}[1]{\ensuremath{\langle #1 |}}
\newcommand{\ket}[1]{\ensuremath{ | #1 \rangle}}
\newcommand{\braket}[2]{\ensuremath{\langle #1 | #2 \rangle}}
\newcommand{\trace}[1]{\ensuremath{{\rm Tr}\{#1\}}}
\newcommand{\nbar}{\ensuremath{\overline{n}}}
\newcommand{\nav}{\ensuremath{\langle n \rangle}}
\newcommand{\gbar}{\ensuremath{\overline{\gamma}}}

\newcommand{\gf}[1]{\ensuremath{g^{(1)}(#1)}}
\newcommand{\gs}[1]{\ensuremath{g^{(2)}(#1)}}

\newcommand{\mattwo}[4]{
\begin{pmatrix}
 #1 & #2 \\
 #3 & #4
\end{pmatrix}
}

\newcommand{\vectwo}[2]{
\begin{pmatrix}
 #1 \\
 #2 
\end{pmatrix}
}

\newcommand{\change}[1]{\textcolor{red}{#1}}

\title{Observation of a Topological Phase Transition in Random Coaxial Cable Structures with Chiral Symmetry}

\author{D. M. Whittaker}
\affiliation{
Physics and Astronomy, School of Mathematical and Physical Sciences,
University of Sheffield, Sheffield S3 7RH, 
United Kingdom.}

\author{Maxine M. McCarthy}
\affiliation{Physics and Astronomy, School of Mathematical and Physical Sciences,
University of Sheffield, Sheffield S3 7RH, United Kingdom.
\\
Max Planck Institute for the Science of Light, 
Staudstra{\ss}e 2, 91058 Erlangen, Germany.
}

\author{Qingqing Duan}
\affiliation{
Wenzhou Institute and Wenzhou Key Laboratory of Biophysics, University of Chinese Academy of Sciences, Wenzhou, Zhejiang 325001, China.
}

\date{\today}

\begin{abstract} 
We report an experimental study of the disordered
Su-Schrieffer-Heeger (SSH) model, implemented in a system of coaxial
cables, whose radio frequency properties map on to the SSH Hamiltonian.
By measuring multiple chains with random hopping terms, we demonstrate
the presence of a topologically protected state, with frequency
variation of less than 0.2\% over the ensemble, evidence of near-perfect
chiral symmetry. When the ends of the chains are connected to form
loops, we observe a topological phase transition, characterised by the
closure of the band gap and the appearance of states which are
delocalised, despite the strong disorder.  
\end{abstract}

\maketitle

\section{INTRODUCTION}

The Su-Schrieffer-Heeger (SSH) model\cite{SSH}, originally a
description of the electronic states in polyacetylene, is one of the
simplest systems of topological physics. It consists of a chain of
sites, representing carbon atoms, connected by hopping terms which
alternate in strength, corresponding to the bonds of the dimerised
molecule. In this periodic form, it has a band gap which closes when
the two hopping strengths are the same. The gap closure separates two
topological phases, determined by the relative magnitudes of the
hopping amplitudes. With appropriate termination, SSH chains can support
localised boundary states which are said to be topologically
protected, because their energy is independent of disorder in the
hopping amplitudes.  These non-trivial topological properties are a
consequence of the chiral, or sublattice, symmetry of the SSH 
model;\cite{table-kitaev,classification-schnyder}
the sites can be divided into two sublattices, such that there are
only hopping terms connecting the two types.

The topology of the SSH model is robust in the presence of disorder,
provided that the chiral symmetry is not broken. A chain with a random
sequence of hopping amplitudes can still be assigned to one of two
topological phases.  We can thus talk about a topological phase
transition in an ensemble of random SSH loops, driven by varying the
parameters in the probability distribution from which the hopping
amplitudes are drawn\cite{random-mondragon-shem}.  { The physics of the
transition in the random chains is much richer than in their periodic
counterpart, a consequence of the larger parameter space.  In the
periodic case, there are only two parameters, the weak and strong
hopping amplitudes, and the phase boundary corresponds to the trivial
uniform chain where these are the same.  For a random chain of length
$N$, the parameter space space is $N$ dimensional, one for each hopping
amplitude, and the structures on the phase boundary are disordered.  }

The theory of such topologically marginal random chains, with
off-diagonal disorder, has a long history, dating back to work by
Dyson\cite{dyson,dyson-theodorou,dyson-eggarter} and continuing through
modern scaling treatments of the Anderson
transition\cite{dos-mckenzie,supersymmetry-balents,scaling-brouwer,fokker-titov,anderson-evers}.
In contrast to the more usual case of on-site disorder, even in one
dimension delocalised states are expected to be found, occurring at zero
energy.  Measurements of localisation properties can thus provide a
signature of a topological phase transition. In an infinite chain, the
transition is also predicted to be accompanied by a singularity in the
density of states.

There have been numerous experimental studies of implementations of
the SSH model using electromagnetic waves, in photonic and microwave
structures\cite{photonic-ozawa,photonic-malkova,microwave-poli,polariton-st-jean,plasmon-bleckmann},
acoustic lattices\cite{acoustic-coutant} and discrete
electronics\cite{electrical-tan,electrical-ningyuan,electrical-lee}. It is generally
hard to control all the couplings in these systems so as to maintain
chiral symmetry with sufficient accuracy to observe the effects we
discuss, particularly while introducing controlled disorder.  
{
As a result,
most demonstrations have focused on periodic structures, without addressing topologically
significant questions about symmetry preserving disorder.} 
However, delocalisation at a topological phase boundary has been
observed\cite{atom-meier} for a momentum-space SSH structure in
cold-atom systems\cite{atom-atala}.

\section{COAXIAL CABLE NETWORKS}

Coaxial cable networks are a very simple electromagnetic system which
can be used to investigate disorder\cite{localisation-zhang} and 
topological\cite{angular-momentum-jiang,non-abelian-guo,cables-oliver,classification-mccarthy} physics.
We have shown\cite{whittaker} that cable structures can be fabricated with radio frequency
properties which map very accurately onto the SSH Hamiltonian. The
hopping amplitudes are determined by the impedances of the
corresponding cables, so it is easy to make a random ensemble of
chains with full chiral symmetry. In this work, we use cable
structures to investigate experimentally the properties of random SSH
chains. By making impedance and transmission measurements, we demonstrate 
very precise topological protection of a state,
and show the delocalisation and closing of the gap at the phase
transition.

The derivation of the matrix description of a coaxial cable network is
given in Appendix \ref{app-tbmatrix}. We consider a network consisting
of a set of sites, labelled $n$, connected by sections of coaxial
cable, all of which have the same transmission time, $\tau$, the length
divided by the transmission speed. The cable connecting sites $n$ and
$n'$ has electrical impedance $Z_{nn'}$. The network has radio
frequency resonances which are determined by a matrix eigenvalue equation $H
v=\varepsilon v$, where the dimensionless `energy', $\varepsilon$, is related to the
frequency, $\omega$, by $\varepsilon=\cos{\omega \tau}$.  The
components of the vector $v$ are the voltages at the sites, scaled
such that the actual voltage is $V_n=\sigma_n v_n$. Here
$\sigma_n=(\sum_{n'} Z^{-1}_{nn'})^{-1/2}$, with the sum taken over
the sites $n'$ directly connected to $n$. The matrix elements of the
`Hamiltonian' are then the hopping amplitudes
\begin{align}
H_{nn'}=\sigma_n Z_{nn'}^{-1} \sigma_{n'}
\;.
\label{eq:hamiltonian}
\end{align}
This one-to-one mapping from cables connecting sites to hopping amplitudes means it is 
possible to create a network corresponding to any finite real matrix Hamiltonian,
though in practice this is limited by the availability of cables with arbitrary impedances.

We measure the radio frequency properties of the cable structures
using a vector network analyser (VNA). 
Two sorts of measurement are
useful.  A single port reflection measurement of the $S_{11}$
parameter gives us the impedance of the structure measured at a given
site. In Appendix \ref{app-tbmatrix}, Eq.(\ref{eq:ldos}), we show that the real part of
this is proportional to the local density of states, broadened only by
losses in the cables. This enables us accurately to determine the
frequencies of the resonances of the structure. Two port
transmission measurements ($S_{21}$) provide information about the spatial
extent of the states, allowing us to detect the delocalisation which
occurs around the phase transition.

\begin{figure}

\begin{center} \mbox{ \includegraphics[scale=0.6]{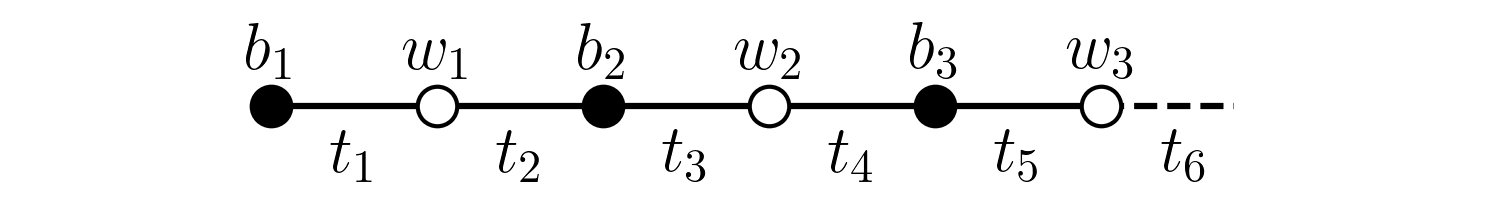} } \end{center}
\caption{
Sublattice colouring of an SSH chain, with black and white sites labelled $b_n$ and
$w_n$, as in Eq.(\ref{eq:Hblock}). The hopping amplitudes, $t_n$, follow Eq.(\ref{eq:pfaffian}).
} 
\label{fig-ssh}
\end{figure}

\section{SPECTRUM AND TOPOLOGICAL PROTECTION}

A Hamiltonian has chiral symmetry if the sites can be divided into two 
sublattices, which we call `black' and `white', Fig.\ref{fig-ssh}, such that there is only hopping
between sites on different sublattices. There can be no intra-sublattice terms,
including on-site energies. If this is the case, ordering the basis such that
all the black sites precede the white sites gives a block anti-diagonal form:
\begin{align}
H\vectwo{b}{w}=\mattwo{0}{Q}{Q^\dagger}{0} \vectwo{b}{w}=\varepsilon
\vectwo{b}{w}\;,
\label{eq:Hblock}
\end{align} 
where $b$ and $w$ are vectors of the amplitudes on the black and white sites. 
From this we obtain $(Q^\dagger Q) w= \varepsilon^2 w$, so the 
eigenvalues  must either be zero, or occur in symmetric
pairs with opposite signs.  It immediately follows that for a chain
with an odd number of sites there must be at least one zero energy
state. Since this conclusion does not depend on the values of the hopping
amplitudes which form the matrix elements of $Q$, the zero-energy state is topologically 
protected against disorder.
More generally, for a chiral network with $n_b$ black sites and
$n_w$ white sites, there are at least $|n_b-n_w|$ protected states.

\begin{figure}
\begin{center} \mbox{ \includegraphics[scale=0.6]{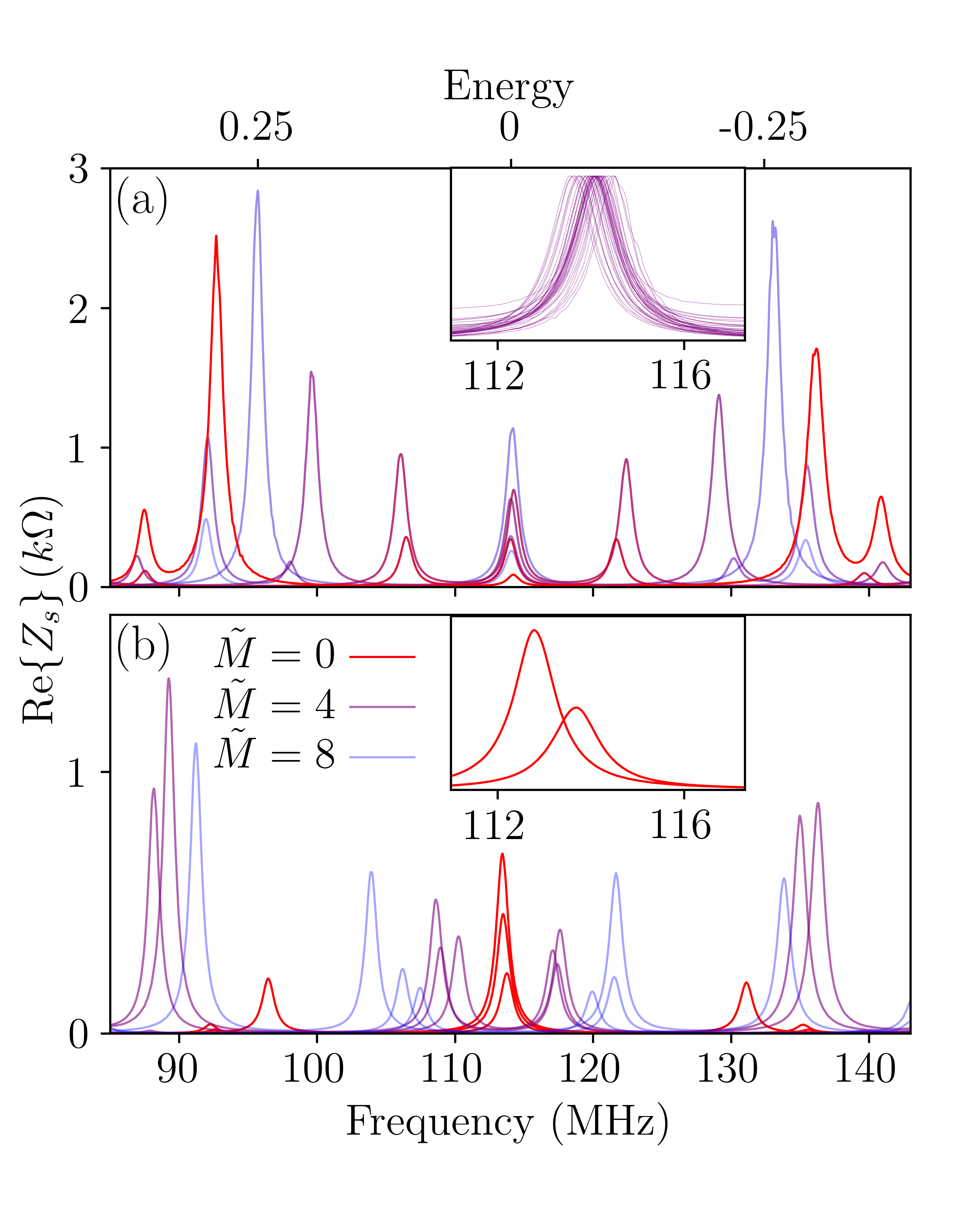} } \end{center}
\vspace{-0.25in}
\caption{
(a) Measured impedance spectra (local density of states) for a selection of unlooped length $N=16$ 
random SSH chains. The colours are an aid to distinguishing curves.
The spectra show the expected symmetry about the chiral frequency ($\sim 114$MHz),
which corresponds to zero  energy in the SSH Hamiltonian. 
The topological protection of the state at $\varepsilon=0$ is apparent. The inset shows, 
expanded and normalised, the protected state in an ensemble of 41 such chains.
The standard deviation of the peak freaquency is 0.22 MHz,  demonstrating the 
minimal chiral symmetry breaking in our cable structures.
(b) Impedance spectra for looped length 16 random cables with various reduced lengths, Eq.(\ref{eq:reduced}):
$\tilde{M}=0$ (red), $\tilde{M}=4$ (purple) and $\tilde{M}=8$ (blue).  Lowering $\tilde{M}$
closes the gap, leading to a doubly degenerate state at the chiral frequency for the
topologically marginal $\tilde{M}=0$ structures.
The inset shows spectra for $\tilde{M}=0$ around  $\varepsilon=0$ measured on adjacent sites, 
revealing the two states, one localised on each sublattice.
Compared to (a),
there is more chiral symmetry breaking, and a slight lowering of the chiral frequency, due to the 
extra length of the T-connector inserted in the loop to make the measurement.
} 
\label{fig-ldos}
\end{figure}

Fig.\ref{fig-ldos} shows the local density of states measurement for a
number of structures consisting of sequences of 16 cables connected end-to-end\cite{orda}.
The individual cables are randomly selected from two impedances: $50\Omega$ and $93\Omega$.
The structures thus map onto finite length SSH chains with randomised hopping terms. 
More details
of the cables are given in Appendix \ref{app-experiment}.
Fig.\ref{fig-ldos}(a) shows spectra for unlooped chains, with the
measurement on a site at the end of the structure. The 16 cables correspond to 17 lattice sites,
so we see, as expected, a topologically protected $\varepsilon=0$ state,
at a frequency of approximately 114MHz. The topological protection is
very good: the inset shows combined results for this state in 41
random structures. The standard deviation of the resonance frequency is
approximately 0.22MHz, which we believe is due to small errors in the
lengths of the cables.

In an infinite periodic SSH structure, a topological phase boundary occurs only for
the trivial situation where the inter- and intra-cell hopping terms
are identical. However, in disordered chains, the possibilities for
marginal structures are more interesting. 
A phase transition is signalled
by the presence of a pair of degenerate states at zero energy,
equivalent to the gap-closure in a periodic structure.  When
$n_b=n_w$, this corresponds to the condition that the determinant
$|Q|=0$\cite{tps-comment}.  For our unlooped structures, $|Q|$ is a simple
product of hopping amplitudes, the first term in
Eq.(\ref{eq:pfaffian}), which cannot be made zero without cutting the
chain, so there is, trivially, just one topological phase.  However,
by joining the ends of the chains to form loops, we can observe a
transition between the two phases of the SSH model.

To have chiral symmetry, the loops must
consist of an even number of sites, and thus be made from an even
number of cables.  In a loop with $N$ cables, labelling hopping
amplitudes rather than the sites, $t_1, t_2, \ldots t_N$
(Fig.\ref{fig-ssh}), we obtain
\begin{align}
|Q|=t_1 t_3 \ldots t_{N-1} - (-1)^{(N/2)} t_2 t_4 \ldots t_N
\;.
\label{eq:pfaffian}
\end{align}
However, since the $t_n$ contain the scaling factors, $\sigma_n$, as well as the cable 
impedances 
$Z_n$, Eq.(\ref{eq:hamiltonian}), it is convenient to look at the  quotient of the two terms,
where these cancel, and define a quantity
\begin{align}
M=2\ln{\left(
\frac{t_2 t_4 \ldots t_{N}}{t_1 t_3 \ldots t_{N-1}}
\right)}
=
2\ln{\left(
\frac{Z_1 Z_3 \ldots Z_{N-1}}{Z_2 Z_4 \ldots Z_N}
\right)}
\label{eq:invariant}
\;.
\end{align}
 For
our structures, where the impedances are taken from a binary distribution, $Z_a$ or $Z_b$,
there are typically cancellations in the ratio of the impedances, and we can write
\begin{align} 
 M=\tilde{M} \ln{(Z_a/Z_b)}
\label{eq:reduced}
\;,
\end{align}
where $\tilde{M}$ is an even integer, which we call the `reduced
length' of the structure.  For a loop where the length $N$ is an even
multiple of 2 ($N=4,8\ldots$) we get $|Q|=0$ when $M$ is zero.  The sign of
${M}$ is thus a topological invariant:  the
impedances, and corresponding hopping amplitudes, are real, so if they were changed
continuously, it would not be possible to flip the sign of $M$ without
passing through a marginal structure with $|Q|=0$. For odd multiples
of 2  ($N=2,6\ldots$) the two terms in the expansion of $|Q|$ have the
same sign, so, though $M$ can be zero, to make a topologically
marginal structure would require a negative hopping amplitude in the
loop. 
Note that the sign of $M$, and thus the attribution of phases, is 
arbitrary, because it is a property of the looped structure and depends on 
the choices of the starting point and direction of numbering. However,
if this is done consistently, the difference in sign distinguishes the two phases.

If, instead of making a loop, the chain is infinitely repeated to form
a periodic structure, we find that the topological classification from
the sign of the reduced length always agrees the generalised Berry
phase\cite{berry-marzari} and winding number invariants obtained from
Brillouin zone based calculations. These methods also predict a phase
transition when $M=0$ in a chain with an odd number of pairs of cables.  In a
periodic structure this is correct, because there are gap closures
somewhere in the Brillouin zone for both even and odd numbers of
pairs. The two cases differ because, for an even number of pairs, the
closure is at wavenumber $k=0$ where the state is the same at the end
of each period, corresponding to the boundary condition for a loop.
For an odd number of pairs, the closure is at $k=\pi$, so the loop
boundary condition is not satisfied.  However, as we show below, the
delocalisation associated with the phase transition can be seen for
both even and odd numbers of pairs. The $M=0$ condition also
corresponds to the phase boundary found in
Ref.\onlinecite{random-mondragon-shem} and observed experimentally in
Ref.\onlinecite{atom-meier}; the unusual reentrant shape of the boundary in these
works is due to the particular choice of rectangular probability distribution
from which the hopping amplitudes are drawn.

In Fig.\ref{fig-ldos}(b), we plot the local density of states for
some random looped chains with length $N=16$ and different values of the
reduced length $\tilde{M}$.  As expected, there is always a gap around
$\varepsilon=0$, except in the marginal case $\tilde{M}=0$, where the
degenerate pair of zero energy states is found.  From this pair, it is
always possible to make states which are localised entirely on
separate sublattices. In the inset, this is demonstrated experimentally by 
comparing spectra from two adjacent sites, one on each sublattice.
The peaks correspond to two distinct states, as can be seen by the small separation in energy.

\section{TRANSMISSION AND LOCALISATION}

In order to explore the localisation of the zero energy states, we make use of transmission
measurements on the unlooped chains. These are most simply described using a transfer matrix treatment,
which relates the currents and voltages entering and leaving the structure. At zero energy,
the transfer matrix, Appendix \ref{app-tbmatrix}, Eq.(\ref{eq:tmat}), for a single cable is
\begin{align}
\vectwo{V_{\rm out}}{I_{\rm out}}=
{\cal M}_n
\vectwo{V_{\rm in}}{I_{\rm in}}
=\mattwo{0}{i Z_n}{i/Z_n}{0}
\vectwo{V_{\rm in}}{I_{\rm in}}
\end{align}
The matrix representing a sequence of $N$ cables is then just the product of the ${\cal M}_n$
for each cable, ${\cal M}={\cal M}_N {\cal M}_{N-1} \ldots {\cal M}_1$. 
The non-zero elements of ${\cal M}$ are the same ratios of impedance products as appear in
$M$, Eq.(\ref{eq:invariant}), so we write, for even $N$, 
\begin{align}
{\cal M}
=(-1)^{(N/2)} \mattwo{e^{-M/2}}
{0}{0}{e^{M/2}}
\label{eq:transfer}
\;.
\end{align}
The measured transmission amplitude, $S_{21}$, at $\varepsilon=0$ is then 
\begin{align}
 S_{21}&=(-1)^{N/2}\mbox{sech}\left(\frac{M}{2}\right)
\nonumber \\
&=(-1)^{N/2}\mbox{sech}\left(\frac{\tilde{M}}{2} 
\ln{\left(\frac{Z_a}{Z_b}\right)} \right)
\label{eq:transmit}
\;.
\end{align}  
For our binary distribution, this is the same as the transmission for
a periodic chain, $Z_a Z_b Z_a Z_b\ldots$,
in which the number of cables is equal to the reduced length
$\tilde{M}$ (for negative $\tilde{M}$ the sequence starts with $Z_b$).
This follows because, at $\varepsilon=0$, the transfer matrix
for an adjacent pair of cables with the same impedance is just minus the unit
matrix, so in calculating $S_{21}$ we can iteratively remove such pairs from the
structure until it is reduced to a periodic chain.

Eq.(\ref{eq:transmit}) shows that the transmission at $\varepsilon=0$
depends only on the value of $M$, and
has a value of unity in chains with $M=0$, which are
topologically marginal when joined to form a loop or repeated
periodically. The topologically protected states in the marginal cables are 
delocalised,\cite{dyson-theodorou,dyson-eggarter} in that they have the same amplitude at either end.  
Away from $M=0$, the state is localised, with a larger amplitude at one or the other end,
depending on the sign of $M$, and thus the topological phase.
The simple treatment leading to Eq.(\ref{eq:transmit}) does not account for the small 
resistive losses occurring in the cables,
though these are easily included numerically.
The losses always reduce the transmission, but they also
cause some spread of the $\varepsilon=0$ values for a given $M$.

\begin{figure}
\begin{center} \mbox{ \includegraphics[scale=0.6]{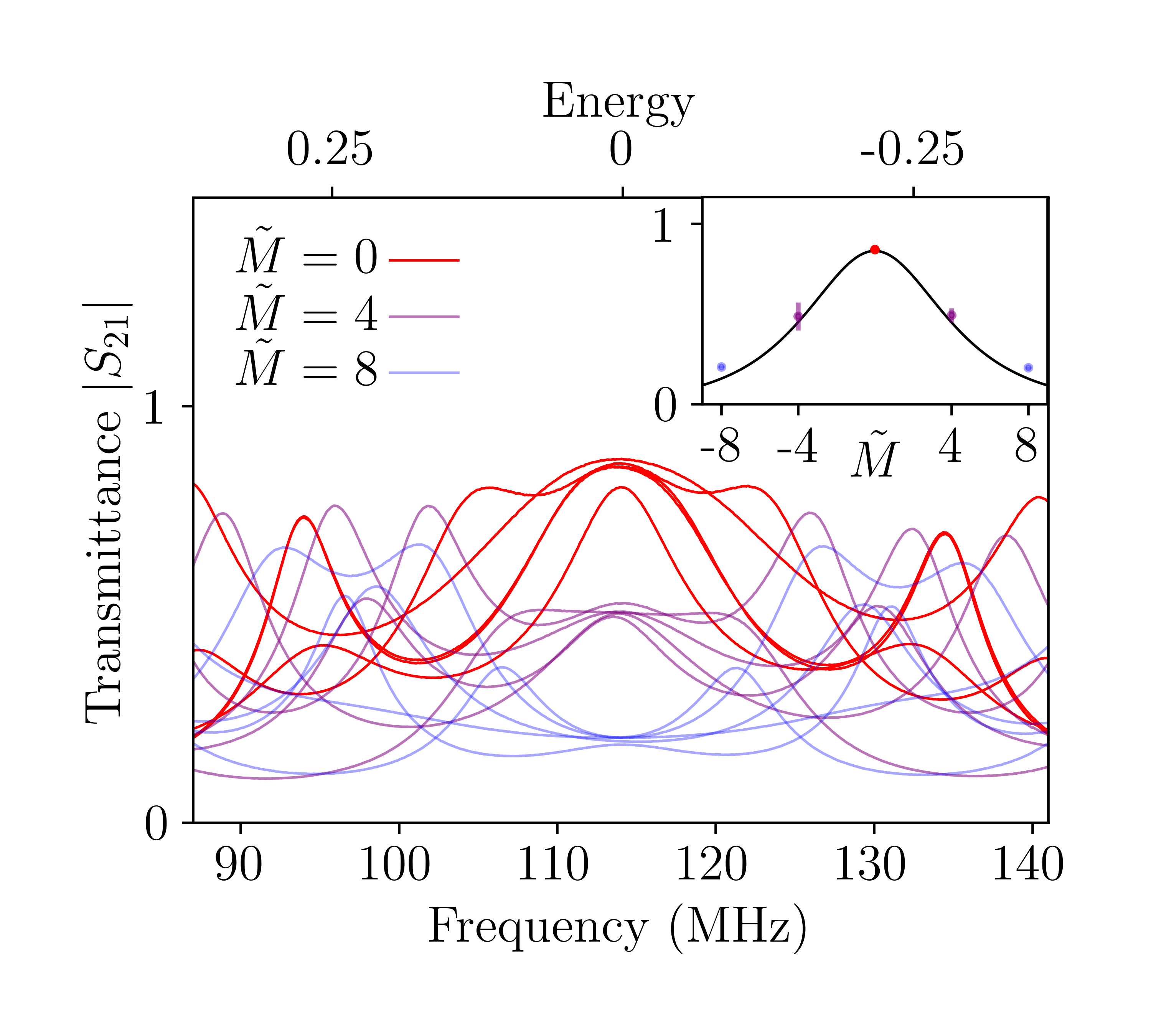} } \end{center}
\vspace{-0.25in}
\caption{
Measured transmission spectra, $|S_{21}|$, for a selection of random unlooped chains with
length $N=16$. The curves are coloured according to the reduced length of the
structures, Eq.(\ref{eq:reduced}): $\tilde{M}=0$ (red), $\tilde{M}=4$ (purple) 
and $\tilde{M}=8$ (blue), as in Fig.\ref{fig-ldos}. The transmission at zero energy (top scale) is, neglecting losses, 
predicted to depend only on $\tilde{M}$. The inset shows the dependence of this transmission
on reduced length (points), averaging over 7 to 9 structures for each $\tilde{M}$.
The error bars show the standard deviation of the measured values.
The solid line
is the behaviour predicted in Eq.(\ref{eq:transmit}), with a constant scaling to account for 
losses.
} 
\label{fig-transmission}
\end{figure}

Experimental transmission results for our length 16 chains are shown in
Fig.\ref{fig-transmission}, where we plot $|S_{21}|$ as a function of
frequency for different values of the reduced length $\tilde{M}$. The
spectra consist of peaks which correspond to the states found in the
$S_{11}$ measurements of Fig.\ref{fig-ldos}, but with much greater
broadening, a result of losses due to the finite ($50\Omega$) input and
output impedances of the VNA.  As predicted, the value at
$\varepsilon=0$ is fairly similar for all structures with the same
$\tilde{M}$. In the inset, the average of $|S_{21}|$ at $\varepsilon=0$
is plotted as a function of $\tilde{M}$, along with the hyperbolic
secant dependence predicted in Eq.(\ref{eq:transmit}), scaled by a
constant factor to account for losses in the cables.  With this scaling,
the agreement is good, and both the absolute values and the spread are
consistent with numerics using the losses deduced from the measured
broadening of the peaks in Fig.\ref{fig-ldos}.

In addition to the delocalisation which we have demonstrated,
Refs.\onlinecite{dyson-theodorou,dyson-eggarter,dos-mckenzie,supersymmetry-balents,scaling-brouwer,fokker-titov,anderson-evers}. predict a singular peak,
the Dyson singularity, in the density of states, $\rho(\varepsilon)$,
around $\varepsilon=0$ for topologically marginal random structures of
infinite length.  This has a functional form $\rho(\varepsilon) \sim |
\varepsilon (\ln{\varepsilon})^3|^{-1}$.  However, in finite
structures this singularity is replaced by a broader peak, which narrows as
$N$ increases. Numerical simulations suggest that, with our choice of impedances,
it is not possible to see such a feature in the short structures we investigate here. 
However, it should be feasible in an ensemble with the lengths increased to 50-100 cables.

\section{DISCUSSION AND CONCLUSIONS}

We have shown that when a looped chiral structure is split, the
transmission through the corresponding chain has unit value, in the
absence of losses, if the original loop was
topologically marginal. Such perfect transmission is thus an
experimental signature of a topological phase boundary. The result
generalises, with some caveats, to more complicated networks with
chiral symmetry. If we start from a balanced structure, having equal
numbers of sites on each sublattice, and break a loop by unplugging a
cable, we split a site, creating an imbalance, and thus a
topologically protected state through which transmission can occur.
The transfer matrix which determines the transmission between the 
the two sides of the break will always be diagonal at zero energy, like Eq.(\ref{eq:transfer}), 
of the form  
\begin{align}
{\cal M}
=\mattwo{\lambda}
{0}{0}{\lambda^{-1}}
\;.
\end{align}
As we have seen, this leads to perfect transmission when $\lambda=1$\cite{transmission-comment}.
However, this is also the condition for the unsplit structure to be topologically marginal; 
then the voltages and currents on either side of the break are identical, which is the boundary 
condition which must be satisfied to obtain a zero energy state when they are 
joined\cite{scattering-fulga}.

The connection between topological phase boundaries and perfect
transmission is not, however, universal. In more complicated networks,
there are cases where a structure is marginal but it can be split in
such a way that the transmission between the ends is less than one,
sometimes zero.  Though a full discussion is beyond the scope of the present
paper, this occurs when, in the split structure, either the
topologically protected state has zero amplitude on the input or
output site, or there is more than one zero energy state on the same
sublattice.

To conclude, we have carried out an experimental study of the
topological properties of a coaxial cable system which maps onto the SSH model.
The accuracy of this mapping is demonstrated by the small variation, less than 0.2\%,
in the frequencies of the topologically protected state in an ensemble
of random structures. By varying the parameters in the random distribution,
we have shown that looped structures can be taken through a topological
phase transition, characterised by the closure of the gap and the appearance
of a delocalised state at zero energy. Coaxial cable structures provide an excellent
platform for such topological physics experiments. Structures with a high degree of
chiral symmetry can be obtained, and disorder introduced in a controlled way, without breaking
the symmetry.  The simple chains considered here can
readily be extended to networks representing more complicated Hamiltonians, where
similar signatures of phase transitions are predicted to be observable.

\begin{acknowledgments}
David Whittaker would like thank Art Gower for many helpful 
lockdown dscussions on this topic.
Maxine McCarthy wishes to thank the EPSRC for a studentship (Project No. 2482664).
Qingqing Duan's work is supported by the National Natural Science Foundation of China 
under Grant No. 12090052. 
\end{acknowledgments}

\section*{Data Availability}

The experimental data used to support the findings of this work are publicly available.\cite{orda}

\appendix

\section{TIGHT BINDING MODEL OF A CABLE NETWORK}
\label{app-tbmatrix}

In this appendix, we derive the transfer matrix for a single coaxial cable and use it
to obtain a generalised eigenvalue equation relating the voltages at the nodes of a network.  
By scaling these voltages, this is turned into a standard eigenvalue problem. 
We then show how to describe the experimental setup by adding driving and loss terms to
represent the ports of the vector network analyser. Finally, we demonstrate the relationship
between the impedance of the network measured at a site and the local density of states.

Our system consists of a series of sections of transmission line of
length $d_n$, with transmission speed $c_n$ and
impedance $Z_n$.  Wave propagation in such a structure 
is determined by the telegraph equations. In each section, the voltage $V(x)$ is
determined by a Helmholtz equation 
\begin{align}
\frac{d^2 V}{dx^2}+\left(\frac{\omega}{c_n}\right)^2 V=0
\;,
\end{align}
where $\omega$ is the frequency. 
At the boundaries between sections both $V$ and the current,
\begin{align}
 I= -i \frac{c_n}{\omega Z_n} \frac{dV}{dx}\;,
\end{align}
are continuous.

For a piece-wise continuous system, we conventionally solve this equation using transfer matrices.
In a given section, the solution can be written
\begin{align}
  V(x) &= V(0) \cos{(\omega x/ c_n)}+ i Z_n I(0) \sin{(\omega x/c_n)}
  \label{eq:voltage}
  \\ 
  I(x) &= iZ_n^{-1} V(0) \sin{(\omega x/ c_n)} + I(0) \cos{(\omega x/c_n)}.
\end{align}  
Thus, at the end of the section $x=d_n$,
\begin{align}
  \vectwo{V(d_n)}{I(d_n)}&= {\cal M}_n \vectwo{V(0)}{I(0)}
\end{align}
where
\begin{align}
  {\cal M}_n =
  \mattwo{\cos{(\omega d_n /c_n)}}{i Z_n \sin{(\omega d_n /c_n)}}
      {i/Z_n \sin{(\omega d_n/ c_n)}}{\cos{(\omega d_n/ c_n)}}
      \label{eq:tmat}
\end{align}

\begin{figure}[h]
\begin{center} \mbox{ \includegraphics[scale=0.6]{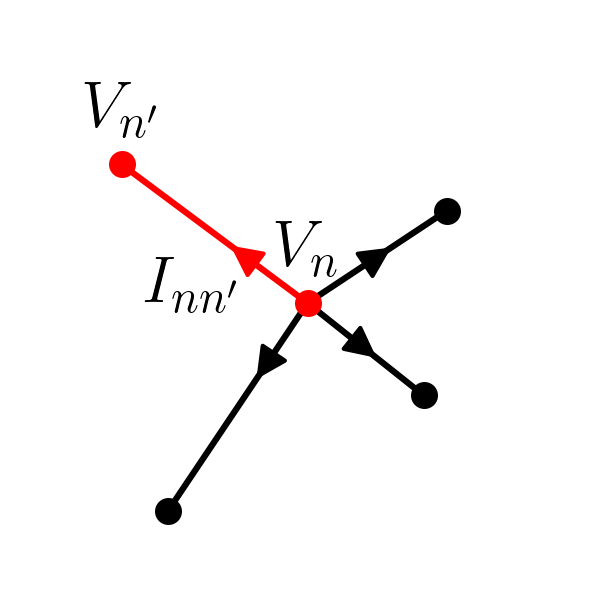} } \end{center}
\vspace{-0.25in}
\caption{
Site $n$ of the network and its neighbours. $I_{nn'}$ is the current flowing out of site $n$
towards $n'$.
}
\label{fig:node}
\end{figure}

We next demonstrate that a transmission line network is equivalent to a tight-binding system when each
section has the same transit time, $\tau=d_n/ c_n$. To show
this, let us add to the notation a little so that $V_n$ is the voltage
at the $n^{\rm th}$ junction (site), and $I_{nn'}$ is the current flowing
out of that junction to connected site $n'$, see Fig.\ref{fig:node}. $Z_{nn'}$
is the impedance of the section between sites $n$ and
$n'$. Then the transfer matrix gives
\begin{align}
  V_{n'}=V_{n}\cos{\omega \tau} + i Z_{nn'} I_{nn'} \sin{\omega \tau},
\end{align}
We can use this to find $I_{nn'}$ in terms of $V_n$ and $V_{n'}$, then
apply Kirchhoff's junction rule,  $\sum_{n'} I_{nn'}=0$, to get
\begin{align}
 \sum_{n'} Z_{nn'}^{-1} (V_{n'}-V_n \cos{\omega \tau})=0\;.
\end{align}
Here, the sum over $n'$ includes the sites which are directly connected to $n$.
Identifying $\varepsilon=\cos{\omega \tau}$, this becomes 
\begin{align}
 \sum_{n'} \, Z_{nn'}^{-1} V_{n'}= \varepsilon \left(\sum_{n'} Z_{nn'}^{-1}\right) \, V_n \;.
 \label{eq:unscaled}
\end{align}
This is a generalised eigenvalue problem, but we can turn it into the standard form by defining
scaled voltages depending on the impedances of the cables connected to site $n$:
\begin{align}
  v_n=\left(\sum_{n'} Z_{nn'}^{-1} \right)^{\!\frac12} \!\!\! V_n = \sigma_n^{-1} \, V_n\;.
  \label{eq:scaling}
\end{align}
Then we have the tight binding system
\begin{align} 
  \sum_{n'} H_{nn'} v_{n'} = \varepsilon \, v_n\;,
  \label{eq:tb}
\end{align} 
with 
\begin{align}
 H_{nn'}=\sigma_n  Z_{nn'}^{-1} \sigma_n' \;.
\end{align}
Note that the `energy', $\varepsilon=\cos{\omega \tau}$, is not the
frequency, so zero energy corresponds to finite frequency; indeed the
spectrum repeats periodically in frequency.

We now consider an experiment where we connect an input to site
$\alpha$, which we model as an ideal voltage source with amplitude $V_{\rm in}$ 
in series with an impedance $Z_{\rm in}$. We measure an output voltage at site $\beta$,
using a detector which is modelled as an ideal voltmeter in parallel with an impedance
$Z_{\rm out}$. The measured output voltage is simply the site voltage $V_{\rm out}=V_{\beta}$.

We start with the detector. From the circuit point of view, this is simply an 
impedance $Z_{\rm out}$, which causes an additional current out of the site $\beta$, 
with magnitude $I_{\rm out}=V_\beta/Z_{\rm out}$. 
The expression from Kirchhoff's rule for site $\beta$ is then modified to
\begin{align}
 \sum_{n'} Z_{\beta n'}^{-1} V_{n'} + i Z_{\rm out}^{-1} \sqrt{1-\varepsilon^2}\, V_\beta=
  \varepsilon \sum_{n'} Z_{\beta n'}^{-1} \, V_\beta \;,
 \label{eq:input}
\end{align}
since $\sin{\omega \tau}=\sqrt{1-\varepsilon^2}$.
Now scaling the voltages as in
Eq.(\ref{eq:scaling}), the tight binding equation for site $\beta$ becomes
\begin{align} 
  \sum_{n'} H_{\beta n'} v_{n'} + i \Gamma_{\rm out} \sqrt{1-\varepsilon^2}\, v_\beta 
  = \varepsilon \, v_\beta\;,
\end{align} 
where $\Gamma_{\rm out}=\sigma_\beta^2 \, Z_{\rm out}^{-1}$; 
the finite impedance leads to an on-site imaginary energy for site $\beta$. 

Doing a similar thing for the source, on site $\alpha$, we get a slightly more complicated result. 
If the source voltage is $V_{\rm in}$, and it has an impedance $Z_{\rm in}$, there is an extra
current flowing into site $\alpha$, $I_{\rm in}$, determined by
\begin{align}
V_\alpha=V_{\rm in} - I_{\rm in} Z_{\rm in}\;.
\end{align} 
Subtracting this current from the Kirchhoff expression gives
\begin{align}
 \sum_{n'} Z_{\alpha n'}^{-1} V_{n'} +  i Z_{\rm in}^{-1} \sqrt{1-\varepsilon^2}\,(V_\alpha-V_{\rm in})=
  \varepsilon \sum_{n'} Z_{\alpha n'}^{-1} \, V_\alpha \;.
 \label{eq:output}
\end{align}
Rearranging, and scaling, we get
\begin{align} 
  \sum_{n'} H_{\alpha n'} {v}_{n'} + i \Gamma_{\rm in} \sqrt{1-\varepsilon^2}\, v_\alpha = 
\varepsilon \, {v}_\alpha + i \sqrt{1-\varepsilon^2}\, {v}_{\rm in}\;,
\end{align} 
where $\Gamma_{\rm in}=\sigma_\alpha^2 \, Z_{\rm in}^{-1}$ and
${v}_{\rm in}=\sigma_\alpha \, Z_{\rm in}^{-1} \,V_{\rm in}$.
The input adds an on-site imaginary energy and a driving term at site $\alpha$.

We now have a matrix system which takes the form
\begin{align}
(H+ i \Gamma) {v} = \varepsilon {v} + i\,{\cal V} \;,
\end{align}
where the `Hamiltonian' $H$ is the same as in Eq(\ref{eq:tb}), and $\Gamma$ is 
diagonal with the only entries the loss terms $\Gamma_{\rm in} \sqrt{1-\varepsilon^2}$ 
and $\Gamma_{\rm out} \sqrt{1-\varepsilon^2}$
on sites $\alpha$ and $\beta$. The driving term
${\cal V}$ has the single entry $\sqrt{1-\varepsilon^2}\, {v}_{\rm in}$ on site $\alpha$.

We can diagonalise $H$ to find its eigenvalues
$\varepsilon_k$, and eigenvectors $u^{(k)}$. Then
$H=U D U^{\dagger}$
where $U$ is the unitary with matrix elements $U_{ij}=u_i^{(j)}$ and $D$ is a diagonal matrix 
with $D_{ii}=\varepsilon_i$. 
This can be used to invert $H- \varepsilon {\mathds 1}$ to get the Green's function
\begin{align}
g=(H- \varepsilon {\mathds 1})^{-1} = U (D- \varepsilon {\mathds 1})^{-1} U^{\dagger},
\end{align}
so that
\begin{align}
g_{ij}=\sum_k \frac{u_i^{(k)} u_j^{(k)}}{\varepsilon_k - \varepsilon}\;.
\label{eq:littleg}
\end{align}
Note that the eigenvectors are real, so the complex conjugate $u_j^{(k)*}$ is
not required.
With this, we can relate the output and input voltages for the case where the input and
output impedances are infinite, so the loss terms,
$\Gamma_{\rm in}$ and $\Gamma_{\rm out}$ are zero:
\begin{align}
{v}_{\rm out}= {v}_\beta 
= i g_{\beta \alpha} \sqrt{1-\varepsilon^2} \, {v}_{\rm in} \;.
\label{eq:glossless}
\end{align}

However, we really want to find $G=(H+i \Gamma - \varepsilon {\mathds 1})^{-1}$, to deal with the 
case where there are finite losses. Since $\Gamma$ has
only two non-zero entries, this can be calculated using the Woodbury matrix identity 
to obtain
\begin{widetext}
\begin{align}
G_{\beta\alpha}=
\frac{g_{\beta\alpha}}
{1+\Gamma_{\rm in} \Gamma_{\rm out} (1-\varepsilon^2)
 (g_{\beta \alpha} g_{\alpha \beta}-g_{\alpha\alpha} g_{\beta\beta})+i \sqrt{1-\varepsilon^2}  
(\Gamma_{\rm in} g_{\alpha\alpha}+ \Gamma_{\rm out} g_{\beta\beta})}
\;.
\label{eq:bigg}
\end{align}
\end{widetext}
Then
\begin{align}
{v}_{\rm out}
= i G_{\beta \alpha} \sqrt{1-\varepsilon^2}\,{v}_{\rm in} \;,
\end{align}
or, in terms of the unscaled physical quantities, 
\begin{align}
{V}_{\rm out}= 
 i \sigma_\beta \sigma_\alpha \, G_{\beta\alpha} \,  Z_{\rm in}^{-1} \,\sqrt{1-\varepsilon^2} \,V_{\rm in} \;.
\label{eq:transmission}
\end{align}

For a single port measurement, connecting only to site $\alpha$ and measuring 
$V_{\rm out}=V_{\alpha}$ to obtain the complex reflectance, we put $\Gamma_{\rm out}=0$ and get
\begin{align}
 \frac{V_{\rm out}}{V_{\rm in}}&=i \sigma_\alpha^2 \, Z_{\rm in}^{-1} \frac{\sqrt{1-\varepsilon^2}\,
 g_{\alpha\alpha}}{1+i \Gamma_{\rm in} \sqrt{1-\varepsilon^2} \, g_{\alpha\alpha}}
\nonumber \\
&=\frac{i \sigma_\alpha^2 \, \sqrt{1-\varepsilon^2}\,g_{\alpha\alpha}}{Z_{\rm in}+i \sigma_\alpha^2 \, \sqrt{1-\varepsilon^2} \,g_{\alpha\alpha}}
\;. 
\label{eq:oneport}
\end{align}
However, we can also think of the circuit as simply a potential divider, with the input 
$V_{\rm in}$ connected across the input impedance $Z_{\rm in}$ in series with an effective 
impedance  $Z_\alpha$ representing the network. In these terms, we get
\begin{align}
\frac{V_{\rm out}}{V_{\rm in}}=\frac{Z_\alpha}{Z_{\rm in}+Z_\alpha}\;.
\end{align}
Comparing to Eq.(\ref{eq:oneport}), the network impedance is thus
\begin{align}
 Z_\alpha=i \sigma_\alpha^2 \, \sqrt{1-\varepsilon^2} \, g_{\alpha\alpha}\;.
 \label{eq:ldos}
\end{align}
Hence
\begin{align}
\frac{{\rm Re}\{Z_\alpha\}}{\pi \sigma_\alpha^{2}\sqrt{1-\varepsilon^2} }=-\frac{1}{\pi}{\rm Im}\{g_{\alpha \alpha}\}=
\sum_k (u_\alpha^{(k)})^2 \, \delta(\varepsilon-\varepsilon_k)\;,  
\end{align}
which is the unbroadened local density of states on site $\alpha$.
In practice, the delta function peaks are broadened by the small resistive losses in the
cables.

{
When there is a single non-degenerate zero-energy state, as for the even length 
open structures considered here, the expression for
$G_{\beta \alpha}$ at zero energy simplifies. Around zero energy, 
$g_{\beta \alpha}$ can be replaced by the divergent term in Eq.(\ref{eq:littleg}), 
$-u^{(0)}_\beta u^{(0)}_{\alpha}/\varepsilon$. Then, $g_{\beta \alpha} g_{\alpha \beta}-g_{\alpha\alpha} g_{\beta\beta}=0$, so
using the expressions for $\Gamma_{\rm in}$
and $\Gamma_{\rm out}$ from above, and putting $Z_{\rm in}=Z_{\rm out}=Z_0$,
\begin{align}
G_{\beta \alpha}(\varepsilon=0)=\frac{-i u_\beta^{(0)} u_\alpha^{(0)} Z_0}
{(\sigma_\alpha u_\alpha^{(0)})^2 + (\sigma_\beta u_\beta^{(0)})^2}
\;.
\end{align}  
This gives the transmission amplitude
\begin{align}
 S_{21}&=2 \frac{V_{\rm out}}{V_{\rm in}}
=2\left(
\frac{\sigma_\alpha}{\sigma_\beta}\frac{u_\alpha^{(0)}}{u_\beta^{(0)}}+
\frac{\sigma_\beta}{\sigma_\alpha}\frac{u_\beta^{(0)}}{u_\alpha^{(0)}}
\right)^{-1}
\nonumber \\
&=2\left(
\frac{V_\alpha^{(0)}}{V_\beta^{(0)}}+
\frac{V_\beta^{(0)}}{V_\alpha^{(0)}}
\right)^{-1},
\label{eq:ratios}
\end{align}
where $V_\alpha^{(0)},V_\beta^{(0)}$ are the physical voltages corresponding to the 
scaled quantities $u_\alpha^{(0)}$ and $u_\beta^{(0)}$. 

For the zero energy state,
Eq.(\ref{eq:unscaled}) simplifies to
\begin{align}
 Z_{n-1}^{-1} V_{n-1}^{(0)}  + Z_n^{-1} V_{n+1}^{(0)}=0\;.
\end{align}
This can be solved iteratively to get, for an even number of cables, $N$,
\begin{align}
\frac{V_{1}^{(0)}}{V_{N+1}^{(0)}}=(-1)^{N/2}
\left(
\frac{Z_1 Z_3 \ldots Z_{N-1}}{Z_2 Z_4 \ldots Z_N}
\right)
\;.
\end{align}
Using the definition of $M$ in Eq.(4), the ratio of the products of odd and even  impedances is 
$e^{M/2}$, so Eq.(\ref{eq:ratios}) recovers the expression for $S_{21}$ obtained from the 
transfer matrix treatment, Eq.(8).
}

\section{EXPERIMENTAL DETAILS}

\label{app-experiment}

We make our experimental structures using two types of coaxial cable:
RG58 and RG62, with impedances of, respectively, 50 and 93$\Omega$. In
order to obtain the mapping onto the tight binding Hamiltonian, and
thus chiral symmetry, it is essential that the transmission time,
$\tau$, in each section of the cable is the same. For our choice of
zero energy at $\sim 114$MHz, this corresponds to nominal cable
lengths of approximately 41cm and 55cm for the RG58 and RG62 cables, as they have
different propagation speeds. To obtain the accurate chiral
symmetry in our results, it was necessary to consider the contribution
of the SMA (SubMiniature version A) connectors used to join the cables, which all have
$50\Omega$ impedance. To account for these, the RG58 cables were
shortened and the RG62 cables lengthened, such that, in a structure
where they alternate, the transmission times in the 50$\Omega$ and
93$\Omega$ sections were the same. However, double sections of the
same cable type are then the wrong length, and the RG62 doubles
contain a pair of 50$\Omega$ connectors in the middle. We avoided these
problems by making double length cables of each type. It is
clearly possible also to make triple and greater lengths, but instead,
we restricted our random sequences to those containing no more than
pair repeats. 

Radio frequency spectra were obtained using a vector network analyser (NanoVNA V2 Plus4).
Our results use two types of measurements. We find the impedance, and thus the local density
of states, using a single port measurement of the $S_{11}$ parameter. The structure impedance is 
then found using
\begin{align}
 Z_\alpha=\frac{1+S_{11}}{1-S_{11}} Z_{\rm in}
\;,
\end{align}
where $Z_{\rm in}$ is the output impedance of the VNA. The value of
the transmission amplitude, $S_{21}$, is obtained directly from a two port
measurement between the ends of the cable.  

The adjustments to the
cable lengths to account for the connectors, as described above, moves
the effective junctions, and hence the sites in the tight binding
model, to the points where the RG62 cables enter their SMA connectors.
In the impedance measurements on the terminated chains, we accounted for
this by calculating the correction to the impedance due to the
transmission through the SMA connector between the physical junction
with the VNA and the effective site position.  At the opposite end to
the measurement, the length of the terminating cable also needed
to be modified, so that the final site corresponds to the end of the
cable. For an RG62 termination, we simply removed the connector at the
end, while in the RG58 case a slightly longer cable accounted for the
pair of SMA connectors required to obtain the correct position.


\begin{thebibliography}{10}
\bibitem{tps-comment}
In more complicated networks, it is also possible to have pairs of topologically protected states
when $n_b=n_w$, in which case $|Q|$ is always zero.
However, this does not occur in the simple structures considered
here.
\bibitem{transmission-comment}
There is also unit transmission  when $\lambda=-1$, which corresponds to a phase transition
when the network is periodically repeated, but not for the finite structure.
\end{thebibliography}
\end{document}